\documentclass[twocolumn,prl,showpacs,floatfix]{revtex4}
\usepackage{graphicx}
\usepackage{amsfonts}
\usepackage{amssymb}
\usepackage{bm}

\newcommand{\be}{\begin{equation}}
\newcommand{\ee}{\end{equation}}
\newcommand{\bea}{\begin{eqnarray}}
\newcommand{\eea}{\end{eqnarray}}

\begin{document}

\title{Numerical Linked-Cluster Approach to Quantum Lattice Models}

\author{Marcos Rigol, Tyler Bryant, and Rajiv R.~P.~Singh}
\affiliation{Department of Physics, University of California, Davis,
CA 95616, USA}

\date{\today}


\begin{abstract}
We present a novel algorithm that allows one to obtain temperature dependent 
properties of quantum lattice models in the thermodynamic limit from exact 
diagonalization of small clusters. Our Numerical Linked Cluster (NLC) approach 
provides a systematic framework to assess finite-size effects
and is valid for any quantum lattice model. Unlike high temperature 
expansions (HTE), which have a finite radius of convergence in inverse temperature, 
these calculations are accurate at all temperatures provided the range of 
correlations is finite. We illustrate the power of our approach studying
spin models on {\it kagom\'e}, triangular, and square lattices. 
\end{abstract}

\maketitle

Understanding finite temperature thermodynamic properties of quantum 
lattice models is a fundamental and challenging task \cite{jaklic,sandvik}. 
Two general approaches that are commonly used are studies of finite 
systems, by means of exact diagonalizations (ED) or quantum Monte Carlo (QMC) 
simulations, and series expansions in the thermodynamic limit (TL). 
ED are usually limited to rather small clusters 
and at finite temperatures and dimensions higher than one it is very 
difficult to assess finite size effects. On the other hand, QMC 
methods enable one to study much larger system sizes 
but then the classes of models that can be addressed are severely 
limited by the (fundamental \cite{troyer05}) sign problem.

In order to obtain results in the TL one can use high temperature 
expansions (HTE). Within this approach properties of the system are 
expanded in powers of inverse temperature, $\beta$ \cite{domb}. 
These expansions, carried out to order $\beta^N$ (where $N$ is 
typically around 10), provide accurate numerical results for 
$\beta<\beta_c$, where $\beta_c$ is the radius of convergence 
of the series. Interestingly, HTE can fail to converge even when 
correlations are still short-ranged. Beyond the region of convergence, 
series extrapolation methods \cite{guttmann} allow one to calculate 
thermodynamic properties, but their reliability remains uncertain.

We introduce in this letter a new method, a Numerical Linked Cluster 
(NLC) approach, that works in the TL as HTE, yet makes possible to 
obtain convergence at all temperatures for models with short-ranged 
correlations. It is also able to deal with multiple microscopic energy 
scales in the problem, which can differ by several orders of magnitude, 
something that is very difficult within HTE.
When the correlation length grows, larger clusters begin to contribute 
and NLC, up to a given cluster size, no longer converges. In some cases, 
one can accelerate the convergence by using sequence extrapolation 
techniques \cite{guttmann,numrecipes}. We will discuss here the 
advantages of NLC over HTE and ED for three different classes of 
models with dominant microscopic energy scale $J$, referred below as:
(A) models that remain short ranged at all temperatures, 
(B) models in which correlations remain short ranged down to
$T<<J$, and
(C) models where correlations build up at $T$ of order $J$.

The fundamental basis for a linked cluster expansion, for some 
extensive property $P$ of an infinite lattice ${\cal L}$, 
is the relation \cite{domb,book}
\begin{equation}
P({\cal L})/N=\sum_c L(c)\times W_P(c),
\label{dirsum}
\end{equation}
Here $N$ is number of lattice sites, $L(c)$ is the lattice constant 
(number of embeddings of the cluster in the lattice per lattice site) 
of cluster $c$, and $W_P(c)$ is the weight of the cluster $c$ for 
the property $P$. The latter is defined recursively
by the principle of inclusion and exclusion \cite{domb},
\begin{equation}
W_P(c)=\mathcal{P}(c)-\sum_{s\subset c}W_P(s).
\end{equation}
Here $\mathcal{P}(c)$ is the property $P$ calculated for the 
finite cluster $c$ and
the sum on $s$ runs over proper subclusters of $c$.
In HTE, for every cluster, $\mathcal{P}$ and equivalently its 
weight $W_P$ is expanded in powers of $\beta$ and only a finite order 
of terms are retained. In NLC an exact diagonalization of the cluster 
is used to calculate $\mathcal{P}$ and hence $W_P$ at any temperature.

Note then that NLC builds in more bare information of the system than HTE. 
We will show that when HTE converges, NLC gives results that 
are identical to HTE. However, NLC converges down to lower temperatures, 
in some cases low enough to obtain ground state properties. In addition, 
unlike HTE, the region of convergence of NLC increases as larger 
clusters are included in the sum. Hence, NLC helps to separate cases 
where the failure of HTE is due to its (not understood) analytic structure 
in the complex plane, from where the correlations truly exceed the largest 
clusters studied.

There is a second aspect in which the NLC scheme is fundamentally 
different to HTE, and that can be used to ones advantage. In HTE, 
the choice of clusters is dictated by the order in which they first 
contribute in the power series, which is typically related to number 
of bonds in a cluster. In NLC, one has substantial freedom to arrange 
the choice of clusters. They can be ordered by number of sites, 
number of bonds, etc. The only requirement is that, with increasing 
order, the cluster weights, when expanded in inverse temperature, 
should give the correct HTE coefficients as well. A small subset of 
the clusters may limit the order to which HTE coefficients are correct. 
Such expansions may sacrifice efficiency in the exact HTE coefficients, 
but can lead to NLC which converge better at intermediate 
and lower temperatures. 

We first apply the NLC method to the {\it kagom\'e} lattice 
Ising model. This model is exactly soluble and known to stay 
disordered at all temperature with a finite entropy at $T=0$. 
Since a {\it kagom\'e} lattice consists of corner 
sharing triangles, it is advantageous to restrict the sum to 
a single site plus clusters that only contain complete triangles,
which reduces dramatically the number of clusters to be considered.
The number of topologically distinct linked clusters on the 
{\it kagom\'e}-lattice with 1 through 8 triangles is 
1, 1, 1, 2, 2, 5, 7 and 15, respectively. (The maximum-site cluster 
with $N$ triangles has $2N+1$ sites.) For the entropy of the 
{\it kagom\'e} lattice Ising antiferromagnet, this leads to a 
rapidly convergent expansion, whose first term is the well known 
Pauling result \cite{liebmann} that gives a ground state entropy 
of $0.50136$. The next correction to this result comes from a 12-site 
cluster of 6 triangles, which leads to $S=0.50182$, that agrees 
with the exact result $S=0.50183$ \cite{exact-kagome} to 4 
significant digits. This very simple example shows that, 
in contrast to HTE, ground state properties of class (A) models 
can be obtained within NLC without the need for extrapolation.  

We now consider the antiferromagnetic Heisenberg-Ising Hamiltonian
\begin{equation}
{\cal H}=\sum_{\langle i,j\rangle} S^z_i S^z_j
+J_{\perp} (S^x_iS^x_j+S^y_iS^y_j)
+h_x\sum_i S^x_i+h_z\sum_iS^z_i,
\end{equation}
where we have chosen the Ising coupling to be unity. 
The transverse-field is denoted $h_x$ and the longitudinal field is 
denoted $h_z$.

As a first application of NLC to a class (A) quantum model, we study 
the {\it kagom\'e} lattice Ising model in a transverse field. 
This model is known to be disordered at all temperatures \cite{msc}. 
In Fig.~1, we show results from the NLC up to 5 and 6 triangle clusters 
for the entropy ($S$) and specific heat ($C_v$) in transverse fields 
$h_x$ of $0$, $0.01$, $0.25$, $0.5$ and $1.0$ respectively. Note that 
the temperature scale goes down to $0.001$, a real challenge for any 
numerical calculation of a thermodynamic system. For $h_x=0.0,0.5$ 
and $1$, the results have fully converged and there is no discernible 
contribution from $6$-triangle clusters at any temperature. Such 
results are beyond the region of convergence of HTE, and in contrast 
to ED of finite clusters they do not suffer from finite size effects.
For $h_x=0.25$, there is a double peaked structure in the specific heat 
and the largest clusters make a small contribution near the lower peak. 
Only for the smallest magnetic field the larger clusters 
contribute. For $h_x=0.01$, the specific heat exhibits two well separated 
peaks. At high temperatures the transverse field plays no role and the results 
are identical to the pure Ising model for the entropy and specific heat. 
Well below $T=0.1$, the transverse field causes the entropy to head down 
towards zero and the second peak arises in the specific heat. 
At the lowest temperatures the correlations are enhanced by the transverse 
field \cite{msc} causing contribution from clusters larger than those 
included in our results.

%%%%%%%%%%%%%%  FIGURE  %%%%%%%%%%%%%%%%%%%%%%%%%%%%%%%%%%%%%%%%%%%%%%
\begin{figure}[!htb]
\begin{center}
  \includegraphics[scale=.58,angle=0] {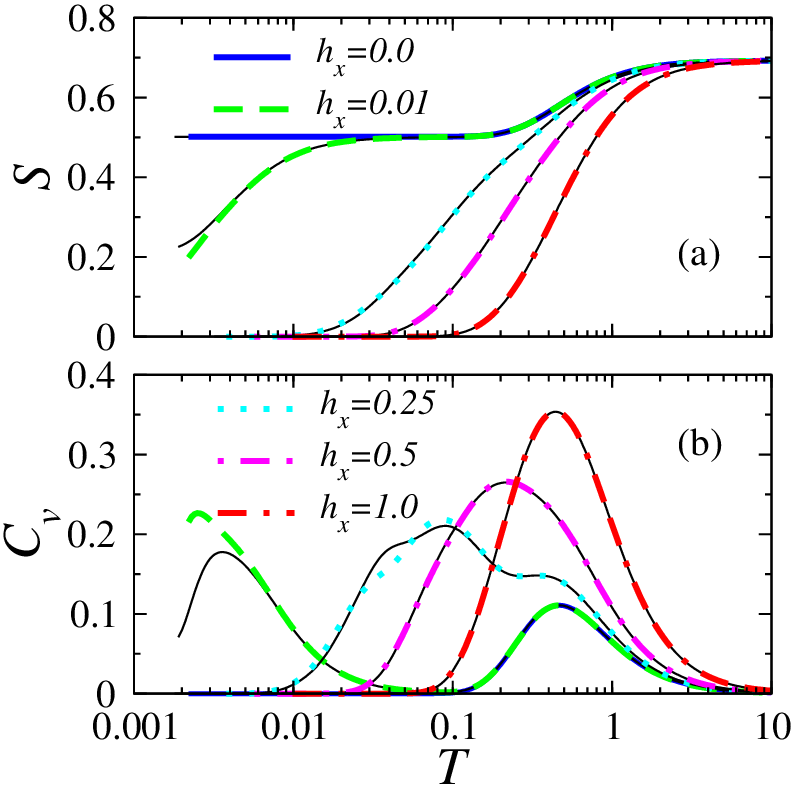}
\end{center}
\vspace{-0.5cm}
\caption{\label{fig1}
(Color online) 
NLC results up to 5 (thin lines) and 6 (thick lines) triangles 
for entropy (a) and specific heat (b) of the transverse Ising 
model on the {\it kagom\'e} lattice as a function of temperature ($T$)
for five different values of the transverse field ($h_x$).
Except for very low but non-zero transverse field, 
the direct sum converges at all temperature.  
}
\end{figure}
%%%%%%%%%%%%%%%%%%%%%%%%%%%%%%%%%%%%%%%%%%%%%%%%%%%%%%%%%%%%%%%%%%%%%%%%

A more challenging, and still open, question is what happens to an 
Ising-like system when quantum fluctuations are introduced via the 
XY coupling ($J_\perp$) [class (B) model]. The {\it kagom\'e}-lattice 
Heisenberg model ($J_\perp=1$) is one of the most fascinating quantum 
spin models, where spin-spin correlations likely remain short-ranged 
down to $T=0$ \cite{finite-size,zeng-elser,singh-huse}. 
Its thermodynamic properties have also been of much interest 
\cite{magnetothermo,elstner,elstner-young}. 
In Fig.~2, we show entropy and specific heat for the XXZ models on 
{\it kagom\'e} lattice with contributions up to 7 and 8 triangles. 
For $J_\perp$ up to around 0.25 our calculations converge down to low 
enough temperatures to see that $C_v$ must have two peaks. As $J_\perp$ 
is increased even further the NLC expansion breaks down before 
a second peak could be resolved \cite{elstner,elstner-young}. The 
temperature dependence is suggestive that a similar ordering mechanism 
is operative for the entire range $0<J_\perp<1$, and quite likely 
large unit cells are involved in further ordering 
at lower temperatures \cite{nikolic}.

%%%%%%%%%%%%%%  FIGURE  %%%%%%%%%%%%%%%%%%%%%%%%%%%%%%%%%%%%%%%%%%%%%%
\begin{figure}[!htb]
\begin{center}
  \includegraphics[scale=.6,angle=0] {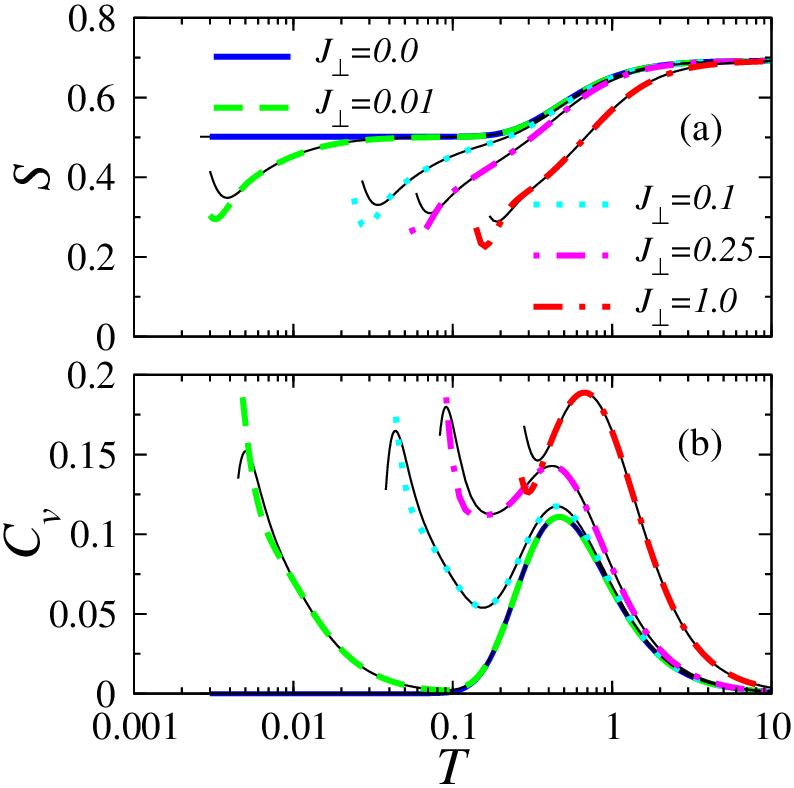}
\end{center}
\vspace{-0.5cm}
\caption{\label{fig2}
(Color online) 
NLC results up to 7 (thin lines) and 8 (thick lines) triangles for the 
entropy (a) and specific heat (b) of the XXZ models on the {\it kagom\'e} 
lattice.
}
\end{figure}
%%%%%%%%%%%%%%%%%%%%%%%%%%%%%%%%%%%%%%%%%%%%%%%%%%%%%%%%%%%%%%%%%%%%%%%%

The high temperature peak in Fig.~2 is associated with short-range order 
and is perfectly resolved whithin our approach. However, it is already 
beyond the radius of convergence of HTE. In Fig.~3 we show a detailed 
comparison between NLC and HTE \cite{elstner-young} for the 
{\it kagom\'e} lattice Heisenberg model. The HTE converges only 
for $T>1$, and inclusion of larger clusters does not improve convergence 
at lower $T$. Only a Pade extrapolation can help HTE at lower temperatures. 
Two such extrapolations from Ref.\ \cite{elstner-young} are also shown.
They can lead to accurate results but their reliability is, in general,
not known. For example, in Fig.~3, one can see that they start to differ
from each other right below the peak in $C_v$. For NLC, on the other hand,
we know that our results are converged if the weight of larger graphs 
is negligible, i.e., it provides a controlled way to approach lower 
temperatures that is somehow absent in HTE and ED. Fig. 3 also shows 
that, as opposed to HTE, the NLC convergence moves to lower temperatures 
as larger clusters are included.

%%%%%%%%%%%%%%  FIGURE  %%%%%%%%%%%%%%%%%%%%%%%%%%%%%%%%%%%%%%%%%%%%%%
\begin{figure}[!htb]
\begin{center}
  \includegraphics[scale=.55,angle=0] {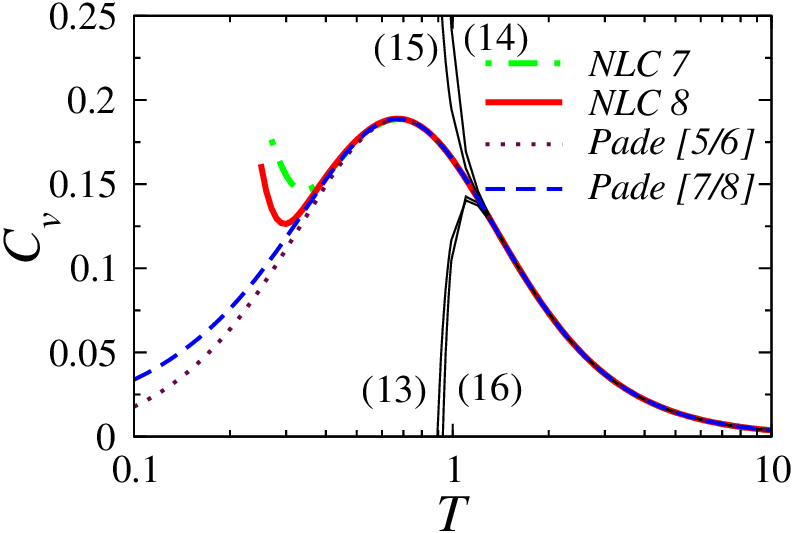}
\end{center}
\vspace{-0.5cm}
\caption{\label{fig3}
(Color online) 
Specific heat of {\it kagom\'e} lattice Heisenberg model as
a function of temperature $T$. The direct sum of
HTE to order 13-16 are shown to diverge around $T=1$. NLC
up to 7 and 8 triangle clusters converges below $T=0.4$.
Two Pade approximants are also shown 
(see Ref.\ \cite{elstner-young} for details), one of 
which is close to NLC result down to $T=0.3$. 
The NLC results indicate
that there may be a second peak below $T=0.3$.
}
\end{figure}
%%%%%%%%%%%%%%%%%%%%%%%%%%%%%%%%%%%%%%%%%%%%%%%%%%%%%%%%%%%%%%%%%%%%%%%%

Within a NLC approach, one way to accelerate the convergence 
of the direct sum in (\ref{dirsum}) is to use sequence extrapolation 
methods \cite{guttmann,numrecipes}. Their power can be seen in Fig.~4, 
where we plot $C_v$ for the Ising model in a transverse and 
longitudinal field \cite{msc} (the phase diagram is shown in the inset). 
With a small transverse field, adding a longitudinal field causes a 
bond-ordered phase to arise. Here we have chosen a small longitudinal field, 
which enhances the correlations of the system, but does not drive it 
into the ordered phase. Larger clusters begin to contribute to specific 
heat below $T=1$, and the simple sum no longer converges at lower 
temperature. However, two different extrapolation methods 
(Euler \cite{numrecipes} and Wynn \cite{guttmann}) lead to results
that converge at all temperatures. Euler's method is a powerful tool 
when terms in the sum alternate in sign \cite{numrecipes}. On the 
other hand, Wynn's algorithm is more general and allows several cycles 
of improvements \cite{guttmann}.

%%%%%%%%%%%%%%  FIGURE  %%%%%%%%%%%%%%%%%%%%%%%%%%%%%%%%%%%%%%%%%%%%%%
\begin{figure}[!hbt]
\begin{center}
  \includegraphics[scale=.55,angle=0] {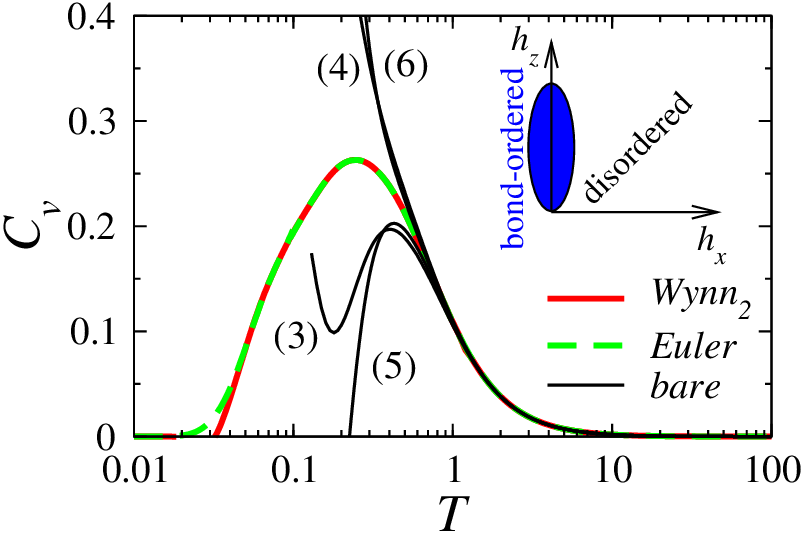}
\end{center}
\vspace{-0.5cm}
\caption{\label{fig4}
(Color online) 
Specific heat of {\it kagom\'e} lattice Ising model in a transverse field 
$h_x=0.5$ and a longitudinal field $h_z=0.25$. The $T=0$ phase diagram 
of the model is shown in the inset (the line $h_x=0$ is critical). 
The parameters correspond to large but finite correlation length, 
where bare sums up to 3, 4, 5 and 6 triangle-clusters diverge below 
$T=1$ but the sequence extrapolation methods converge at all
temperatures. The subindex ``2'' in Wynn means that two cycles were 
applied to accelerate convergence \cite{rigol06_4}.
}
\end{figure}
%%%%%%%%%%%%%%%%%%%%%%%%%%%%%%%%%%%%%%%%%%%%%%%%%%%%%%%%%%%%%%%%%%%%%%%% 

As a final test of our method we study the Heisenberg model on the 
triangular and square lattices [class (C) models], which are known 
to develop long range order at zero temperature. For the triangular 
lattice, we use NLC based on triangular clusters (up to 8 triangles), 
whereas for the square-lattice we use a site-based expansion of up 
to 13 sites. In Fig.~5 we show the entropy of these models obtained 
by various orders of extrapolations with the Wynn and Berezenski 
methods \cite{guttmann}. These results are compared with those 
obtained by Bernu and Misguich (BM) \cite{misguich}. The agreement 
is quite good down to $T=0.3$ for the square-lattice case, where 
the entropy is $<0.05$ (spin-spin correlation length about 20 lattice 
spacings \cite{elstner}, i.e. larger than our cluster sizes), whereas 
for the triangular-lattice it is good down to $T=0.2$ where the
entropy is about $0.2$ (spin-spin correlation length about two lattice 
spacings \cite{elstner}, which raises the question, what other 
correlations are building up leading to a breakdown of NLC convergence?). 
In general, the extrapolations do not converge well below the 
peak for the specific heat. Hence, a priori, 
there is no advantage of NLC (with extrapolation) over HTE 
(with Pade extrapolations) for models of class (C), as both methods 
require extrapolations, whose convergence is difficult to judge.
NLC, however, does provide a scheme that like HTE allows for 
systematic extrapolations.

%%%%%%%%%%%%%%  FIGURE  %%%%%%%%%%%%%%%%%%%%%%%%%%%%%%%%%%%%%%%%%%%%%%
\begin{figure}[!htb]
\begin{center}
  \includegraphics[scale=.55,angle=0] {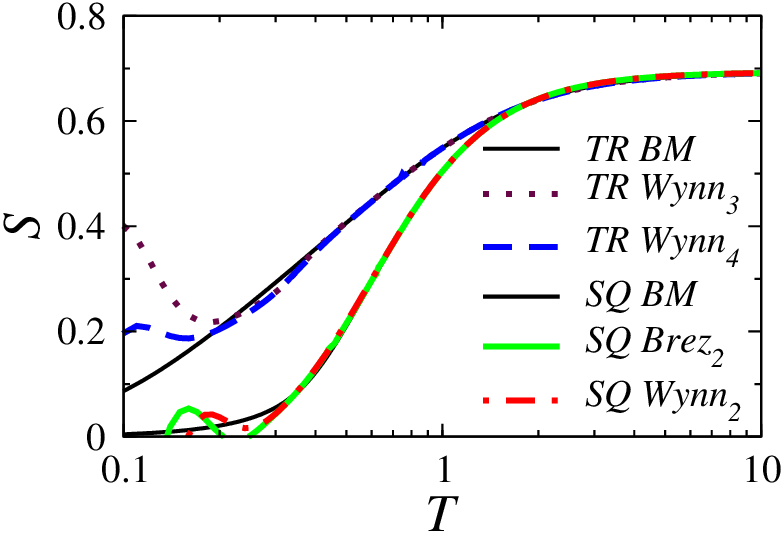}
\end{center}
\vspace{-0.5cm}
\caption{\label{fig5}
(Color online) 
The entropy of square (SQ) and triangular (TR) lattice Heisenberg
models. Various extrapolations of NLC are compared with results
obtained by Bernu and Misguich (BM) \cite{misguich}. Subindexes 
in Wynn's and Brezinski's (Brez) extrapolation results show the number
of cycles applied to accelerate convergence \cite{guttmann}.
}
\end{figure}
%%%%%%%%%%%%%%%%%%%%%%%%%%%%%%%%%%%%%%%%%%%%%%%%%%%%%%%%%%%%%%%%%%%%%%%%

In summary, we have introduced a Numerical Linked Cluster method
to calculate properties of quantum lattice models. It provides
a framework to study observables in the TL while performing exact 
diagonalization on finite-size clusters. This approach allows us 
to go beyond the radius of convergence of HTE, and is better suited 
to models where correlations remain short-ranged down to low temperatures.

We have used NLC to study the thermodynamic properties of frustrated
two-dimensional quantum antiferromagnets. We have shown that one can 
obtain accurate results for short ranged models (Ising and transverse 
Ising models on a {\it kagom\'e} lattice) at all temperatures.
For models where correlations develop slowly, like XXZ models
on the {\it kagom\'e}-lattice, there is a large temperature window 
(which is $0.3<T<1.0$ for the Heisenberg case) where NLC converges 
but HTE diverges. Hence, NLC provides for these [class (B)] models a 
framework to approach lower temperatures in a controlled way. 
We have also shown that the region of convergence 
of NLC increases as larger clusters are included. 

In order to accelerate the convergence of NLC one can use 
sequence extrapolation methods, but then uncertainty similar 
to Pade extrapolations for HTE remains. As examples we have studied
Heisenberg models on triangular and square lattices, where our 
results compare very well with those in Ref.\ \cite{misguich}. 
To study models of interest at lower temperatures our approach can 
be extended to include larger cluster by using Lanczos type methods 
focussing only on low lying states rather than a complete 
diagonalization \cite{jaklic}. Furthermore, the method can be 
applied to $t-J$ and other models, and to various susceptibilities 
and correlation functions. These are left for future work.

\begin{acknowledgments}

This work was supported by the US National Science Foundation, 
Grant No. DMR-0240918 and DMR-0312261. We are grateful to 
B. Bernu and G. Misguich for providing us with their data from 
Ref.\ \cite{misguich}.

\end{acknowledgments}

\end{document}